\documentstyle[editedvolume,psfig]{crck}

\begin{opening}
\title{Astronomy in Ukraine}
\subtitle{Status of Astronomical Research} 
\author{Ya.V.Pavlenko}
\institute{Main Astronomical Observatory of the National Academy of Sciences,
           27 Zabolotnoho, Kyiv-127, 03680 Ukraine}
\institute{Centre for Astrophysics Research, University of Hertfordshire,
College Lane, Hatfield, Hertfordshiere AL10 9AB, UK}
\author{I.B.Vavilova}
\institute{Dobrov Center for Scientific-Technical Potential and History of 
Science Research, National Academy of Sciences of Ukraine
60 T.Shevchenko Blvd., Kyiv 01032 Ukraine}
\author{T. Kostiuk}
\institute{NASA/Goddard Space Flight Center,
Greenbelt, MD 20771, USA}

\end{opening}

\runningtitle{Astronomy in Ukraine}

\begin{document}


\begin{abstract}
The current and prospective status of astronomical research in Ukraine
is discussed. A brief history of astronomical research in Ukraine is
presented and the system organizing scientific activity is described,
including astronomy education, institutions and staff, awarding higher
degrees/titles, government involvement, budgetary investments and
international cooperation. Individuals contributing significantly to
the field of astronomy and their accomplishments are mentioned. Major
astronomical facilities, their capabilities, and their instrumentation
are described. In terms of the number of institutions and personnel
engaged in astronomy, and of past accomplishments, Ukraine ranks among
major nations of Europe. Current difficulties associated with political,
economic and technological changes are addressed and goals for future
research activities presented.
\end{abstract}

\section{Introduction}  
Since resuming its independence in 1991 Ukraine has been striving 
towards social democracy and a market-based economy. In 2005 
after the ``Orange Revolution'' these goals as well as European 
integration became priorities of Ukraine's policy. Science, and 
in particular astronomy, also has to adopt a model that matches the 
emerging market-based economy. Recently Ukrainian science has 
experienced complicated institutional and structural changes in 
the state-administered system formed during Soviet times. The basics 
principles of the reforms are: to match scientific endeavors to 
the economic capabilities of the nation; to form a mechanisn to 
address legal and economic issues, including protection of 
intellectual rights and appropriate coordinatation and budgeting 
for effective governing of the scientific and technological field; 
to specify science and engineering development priorities; to 
radically improve the resource management and to address the 
effects of aging and the shortage of personnel; to follow the 
principle of ``openness'' in science; and to promote wide international 
cooperation (Yatskiv \& Vavilova 2003). 

\subsection{Geographical location} 
Ukraine is one of the largest countries on the 
European continent covering 603,700 sq. km in area. Its territory 
stretches 1316 km east-west and 893~km north-south. Ukraine borders 
Poland, Slovakia, Hungary, and Romania in the west, Moldova in the southwest, 
Belarus in the north, and Russia in the northeast and east (Fig. 1). 
The southern frontiers of Ukraine are washed by the Black Sea and the 
Sea of Azov with a coastline equal to 2835~km.

\begin{figure}
\begin{center}
\psfig {file=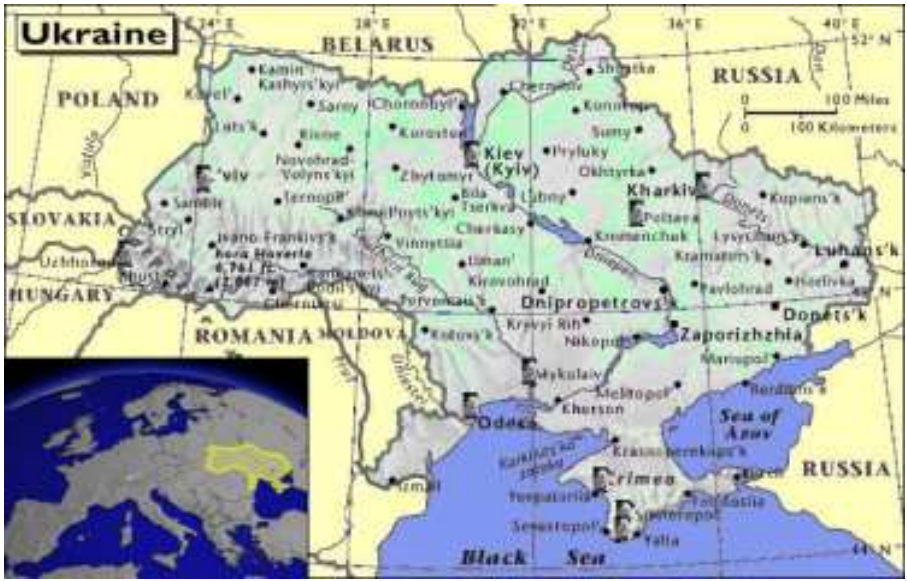,width=120mm}
\end{center}
\caption{Astronomical sites in Ukraine}
\end{figure}

\subsection{Population} 
Ukraine's population on October 1, 2004 was 47 383 486 \\ 
(www.ukrcensus.gov.ua/results). The population is ethnically diverse 
with 77.8\% ethnic Ukrainian. The urban population is about 67.2\% and 
the population density is about 80 inhabitants per sq. km.

\subsection{Political System and Government} 
Ukraine has a presidential political 
system with a one house parliament (Verkhovna Rada). Some changes 
strengthening the parliament were introduced during the ``Orange Revolution'' 
but are scheduled to come into effect in 2006. 

The capital is Kyiv (Kiev) with 2 611 000 inhabitants. The state 
administrative system consists of the Autonomic Republic of Crimea 
with Simferopol its capital, 24 regions (i.e., oblasts), 481 
administrative-territorial districts. The cities of Kyiv and 
Sevastopol (Crimea) have a special administrative status. There 
are 451 cities, 893 towns, and 28651 villages in the country. 

\subsection{Education} Ukraine is a highly-educated country where 28 900 000 
citizens have received a secondary or higher education. In the age 
group of 18 years and older, 13 700 000 persons have some form of 
higher education \\ (www.ukrcensus.gov.ua).). It also is important to 
note that the full legal age in Ukraine is 18 years and coincides 
with the age when a citizen completes his/her secondary education. 

At present, there are 48,000 educational institutions, where 10 million 
scholars and students are being trained by 1.5 million teachers and lecturers. 
Among them, 2 700 000 citizens (of which 54\% are females) are students 
of 966 higher educational institutions. There, 178 000 lecturers are 
engaged in the educational process -- 44 500 have a Candidate of Science 
degree, 7 400 have a Doctors of Science degree, among them 7 500 have the 
academic status of Professor and 31 600 of Assistant Professor 
(Nikolaenko 2004). Two hundred higher educational institutions have 
University status or National University status. At present, the Ministry 
of Education and Science of Ukraine, MESU, which is responsible for 
development of education, is working to improve the efficiency of 
this system in the framework of the Bologna Process   
(http://www.aic.lv/ace/ace\_disk/Bologna/index.htm or 
http://www.pfmb.uni-mb.si/bologna/Joint\%20declaration\%20of\%20the \\
\%20European\%20Ministers\%20of.htm). 
This Process aims to standardize the various European higher education 
systems with the objective of creating a European Area of Higher Education 
and of promoting the European higher education system worldwide. 
On May 19, 2005 Ukraine joined the Bologna Process. 

\subsection{Science} 
There are three governmental structures which are mostly 
responsible for development of science and technology (S\&T): 
the National Academy of Sciences of Ukraine (NASU), 
the Ministry of Industrial Policy, and the Ministry for Education 
and Science. Beginning in 1991, the total expenditure 
(budgetary and off-budgetary) for research and development (R\&D) 
has been reduced by factor of 4 (750 000 000 USD in 2004). The 
gross expenditure on R\&D as a percentage of GDP also has been reduced 
by a factor of two. The total budgetary expenditure on S\&T
 in 1991 -- 2002 in relative terms of purchasing-power 
parity has also deteriorated by one-half to 2 030 000 USD in 2002 
(Yatskiv 2004). 

Science in Ukraine, including astronomical research, is now facing a 
difficult time due to economic limitations of the nation and the need 
for up-grading the existing scientific infrastructure. The key problems 
are both the low GDP activities and the fact that even the low budgetary 
expenditures on science and technology have not been effectively spent. 
In 2005 the adopted budget is 20 400 000 000 USD with 5\% devoted to R\&D. 

In 2004 about 173 000 employees were engaged in science and technology 
activity: 54\% of them worked in industry, 28\% in national academies of 
science, and 18\% in higher educational institutions. In this group 73 700 
persons are Candidates and Doctors of Science As was mentioned above, 
70\% of them work in higher educational institutions. If one considers 
the number of researchers per 10 000 labor force, we see that Ukraine with 
41 is at the level of countries such as, Germany -- 58, Great Britain -- 54, 
Austria -- 34.

Beginning in 1991, there has been an increase in the number 
of post-graduate students (13 600 in 1991 and 24 500 in 2003). Although the 
total number of scientists has decreased during this period because of 
a ``brain drain'' both outside and inside Ukraine, the number of Doctors 
and Candidates of Sciences has been slowly increasing. A greater fraction 
of those taking science and engineering degrees during this period 
were women: 54\% of graduate students, 54\% of Candidates of Science, 
27\% of Doctors of Science, and 10\% of members of academies of sciences. 

\section{ Brief History of Astronomical Research}

Astronomical culture and research have long-standing traditions in the 
country. The first signs of astronomical knowledge were found in 
archaeological excavations and records. The most ancient find (dated as 
15~000~B.C.) is a mammoth tusk with a fretwork image of a table of lunar 
phases found in the Poltava region. The so-called Trypillya culture 
(dated 4 000 -- 3 000~B.C) had numerous examples of  ornaments at the 
howls, distaffs, wheels and other everyday articles with symbolic images 
of zodiac constellations, and vessel-calendars indicating the 
vernal/autumnal equinoxes and the motion of the Sun.

Another unique historical record relates to the times of the powerful 
state of the Kievan Rus' (X -- XIII centuries), when astronomical 
observations were conducted mainly in cloisters. For example, the authors 
of the Lavrentievska chronicle describe the solar eclipses of the years
 1064, 1091, and 1115 A.D. and the lunar eclipses of 1161 A.D.

The first book on astronomy written in the modern territory of Ukraine 
was by physician and astronomer Georgii Drohobich in 1483, who later 
became the Rector of Bologna University 
(http://litopys.org.ua/human /hum47.htm). A graduate of the Kyiv-Mohyla  
Academy, Ivan Kopievsky, issued the first stellar map in the Slavic 
language in Amsterdam in 1699 and the basics of naval astronomy in 1701. 
The prominent Ukrainian-Russian philosopher, scientist and religious
figure, Pheophan Prokopovich, who worked at the Kyiv-Mohyla Academy in
1705 -- 1716 and was rector in 1711 -- 1716, lectured astronomical courses
based on theories of Copernicus and Galileo. He also developed the
philosophical foundations of the unity of matter and motion, which were
generalized later by Mikhail Lomonosov.

The first astronomical observatory on the territory of western Ukraine was 
founded in L'viv University in 1769. The first astronomical observatory in 
Kyiv was founded in the library of the Kyiv-Mohyla Academy in 1783 
and was equipped with modern instruments of those times. 
In 1821 the Naval Observatory was established in Mykolaiv, which
functioned in 1912-1991 as a department of the Pulkovo Observatory (St.
Petersburg, Russia). 
Later on university observatories were founded 
in Kyiv (1845), Odesa (1871), and Kharkiv (1883). 
In 1909 a southern department of the Pulkovo observatory was established in
Simeiz (Crimea). 
Being engaged in the educational process, these observatories conducted 
research in astrometry, theoretical astronomy, and astrophotometry.

The Ukrainian Academy of Sciences was founded in 1918. This resulted in 
a new impetus for the development of science and technology, in 
particular, the establishment of new astronomical institutions and 
infrastructure. Among them are the Gravimetric Observatory in Poltava 
(1926), which is now a division of the Institute of Geophysics; the 
Main Astronomical Observatory, MAO, in Kyiv (1944); the Crimean 
Astrophysical Observatory, CrAO, in Naukove, Crimea (1945); the Radio 
Astronomical Observatory of the Institute of Radio Physics and Electronics 
in Grakovo (1958), which later became a division of the Institute of 
Radio Astronomy, IRA, in Kharkiv (1985); the High-Altitude Observatory 
at the Peak Terskol, the North Caucasus  (1970's) which later became a 
division of the International Center for Astronomical and 
Medical-Ecological Research, ICAMER (1992). Currently, the astronomy 
sector of the National Academy of Sciences of Ukraine is well-developed 
as compared to that at the universities.

The turning points in the history of science are inseparably 
linked with outstanding personalities. Among them are the founders 
of modern astronomy in Ukraine who obtained world recognized 
achievements in the years 1920 - 1970: A. Y. Orlov and E.P.Fedorov 
(astrometry, complex research of the Earth rotation); 
N. P. Barabashov (photometry of planets, the first catalogue of 
details of the Lunar surface, developer of one of the first 
spectrohelioscopes for research of the solar photosphere and 
chromosphere); G.A.Shajn (stellar evolution, kinematics and 
the magnetic field of the Galaxy, solar corona, long-periodic 
variable stars, discovery and research of new emission nebulae); 
A.B.Severnyj (spectral research of solar flares and other non-stationary 
processes on the Sun, magnetic fields of the Sun and stars, building 
of one of the largest solar telescopes, the Tower Solar Telescope, 
solar oscillations, helioseismology, the problem of solar  
deuterium, development of space-born telescopes and research 
programs for space missions); V.P.~Tsesevich (complex research of 
variable stars); A.Y.Yakovkin (research of the Moon); 
S.K.Vsehsvyatskij (cometary research, theories on volcanic activity 
on satellites and the existence of rings around Jupiter, 
independent discovery of the solar wind altogether with Ponomariov); 
S.A.Kaplan (general relativity theory, white dwarfs, 
gas-dynamical processes); S.Ya.Braude, one of the founders of 
decameter radio astronomy (developer/builder of the largest decameter 
telescope UTR-2 and the decameter VLBI URAN system, author of the 
first catalogue of extragalactic objects in the decameter radio range).

\section{The Current Status of Astronomical Research} 

\subsection{Secondary And Higher Astronomical Education}

\subsubsection{Secondary Education} Without exaggeration we can say that the system 
of education and its achievements at this level were the most 
extraordinary accomplishments of the former Soviet Union, 
in particular as it concerned education in astronomy. Astronomy 
was a basic course in secondary schools (34 academic hours in the last, 
10th, grade). Nobody knows the reasons, although it is possible it was 
an echo of reform, but after Ukraine resumed its independence in 1991, 
from 1992 until 2000 astronomy was excluded from the secondary education 
basic curriculum. In 2000, as a result of persistent activity by the 
Ukrainian Astronomical Association (UAA) and numerous round-tables with 
representatives of ministry departments, this regretable decision was 
corrected and astronomy was reinstated into the current 12-year secondary 
education curriculum. 
     
The present-day status of the astronomical education in secondary schools 
is as follows (Vavilova \& Yatskiv 2003):  
	
Some elements of the astronomical discipline are included in the standard 
``Natural Science'' curricullum of the 6th - 11th grades.

``Astronomy'' is a required course in general (non-specialized) schools 
(17 academic hours in the last 12th grade) and in lyceums of the natural 
sciences (34 academic hours in the 12th grade). 

``Astronomy'' as an ellective course is studied in gymnasiums of the 
humanities.

By comparing the status of secondary astronomical education in 
other countries (e.g., Russia, several countries of Europe) the UAA timely
improved this situation in Ukraine by arguing that knowledge of astronomy 
will play a unique role for generations to come in the 
twenty-first century. Our present-day problems in this improvement are 
how to increase the number of new textbooks and how to organize regular 
training of astronomy teachers. 

Several planetariums are open to the public. Two of them, in 
Kyiv and Kharkiv, are located in separate buildings. The Drahomanov 
Pedagogical University of Kyiv, the Pedagogical University of Mykolaiv, 
the National University of Uzhgorod also train teachers of astronomy 
for secondary schools. 

\subsubsection{Higher astronomical education} We list below the most 
important national 
universities, which have astronomy and space related faculties:

\begin{itemize}

\item	Shevchenko National University of Kyiv (www.univ.kiev.ua)
\item	V.N. Karazin National University of Kharkiv (www.univer.kharkov.ua)
\item	I.I. Mechnikov National University of Odesa (www.onu.edu.ua)
\item	Ivan Franko National University of L'viv (www.franko.lviv.ua)
\item	National Technical University ``Kyiv Polytechnical Institute'' \\ 
(www.ntu-kpi.kiev.ua ) 
\item	National University of Dnipropetrovsk (www.dsu.dp.ua) 
\item	V.I. Vernadsky Taurian National University in Simferopol \\ 
(www.ccssu.crimea.ua/tnu) 
\item	National University of Uzhgorod  (www.univ.uzhgorod.ua) 
\item	Zhukovsky National Aerospace University in Kharkiv (www.xai.edu.ua)

\end{itemize}
 
All astronomical programs are structural divisions of the physics 
departments of universities. For this reason as well as the fact that 
the Ukrainian system of university education in the natural sciences is 
similar to that of the German system, our students-astronomers receive 
good training in mathematics and physics.     

Every year a total of about 75 university entrants are educated in 
astronomy. After the 4th year they obtain a Bachelor diploma in Physics 
and on graduating from the university they obtain either a Diploma 
of Specialist or a Master's Degree in Astrophysics/Astronomy. They 
study the classical university courses (astrometry, celestial mechanics, 
planetary physics, solar physics, astrophysics, applied astrophysics, 
theoretical astrophysics, extragalactic astronomy etc.) as well as 
special courses on contemporary astronomical research, and have seminars 
and training in observational astronomy. Results of our monitoring show 
that 80 \% of the entering students finish their education in 5 years; 
50 \% of students, who finished their education, continue to work in 
astronomy; 30\% of holders of a Specialist's diploma or Master's Degree 
defend a Candidate Thesis within 3 - 7 years after they graduate 
(Vavilova \& Yatskiv 2003). 
 
Current problems in implementation of the astronomical education program 
are the following: 

\begin{itemize} 

\item	There are no required astronomical courses in the first 
	to third year of education in the bachelor programs (only 
	classical courses for physicists)
\item	Many textbooks need to be renewed
\item	There is a drop of the mean level in the ability of our students 
during the last few years and an indication of this problem is also evident 
in secondary education 
\item	The need to raise the prestige of the scientific profession 
irrespective of the low salaries for young scientists and engineers.

\end{itemize}

\subsection{Scientific Degrees and Scientific Careers of Astronomers}

\subsubsection{Scientific degrees} 
The system of Bachelor and Master's degrees was initiated in 2002. 
For this reason we are not ready yet to analyze how it works. The system 
of higher scientific degrees in Ukraine is inherited from the Soviet type 
system and consists of two levels: Candidate of Science, 
and Doctor of Science. During recent years the Ministry of Foreign 
Affairs of Ukraine approved memorandums about compliance of diplomas 
of Candidate of Science and PhD with dergrees of more then 100 
countries of the world. Everybody with a degree up to the Cand. 
Sci. Diploma may obtain a certificate of compliance for work abroad.

The first level degree of Candidate of Sciences (Cand. Sci.) 
is the analogue of the degree of Doctor of Philosophy or Doctor of 
Medicine adopted in Europe and other countries. 
The Cand. Sci. degrees unlike the Ph.D. degrees are classified by 
the related scientific fields (chemistry, biology, pedagogy, 
economy, politics etc.). The second level degree of Doctor of 
Science (Dr. Sci.) is also classified by related scientific fields. 
The Degrees of Cand.Sci. and Dr.Sci. for those who work in astronomy 
are related to such fields as the Physics-Mathematical Sciences and 
Technical Sciences. 

As a rule, the Cand. Sci. thesis in astronomy is a manuscript (90 --  120 
pages) based on at least 3 papers published in refereed journals and 
resulting in a new achievement in astronomy.  The Dr.Sci. thesis is 
a manuscript (220 -- 260 pages) based on: a) significant scientific 
discovery in Astronomy; b) advanced results published in at least 
20 papers in refereed journals; c) an author's monograph.

The thesis of Cand. Sci. and Dr. Sci. is defended following a special 
procedure. The first step is a report at a Scientific Council meeting 
of the institution where the thesis has been conducted. This Scientific 
Council recommends/does not recommend a thesis defense. The second step 
is a report at the meeting of the Special Council on Thesis Defense, 
which nominates/does not nominate the candidate for a Cand.Sci. or Dr.Sci. 
degree as well as decides whether the thesis satisfies the requirements 
of a Cand. Sci. or Dr. Sci. thesis. Candidates for a Cand. Sci. degree 
must also take exams in philosophy, foreign language, and in their 
specialty before the defense.

For defense of the Cand. Sci. thesis the Special Council selects 
two reviewers, 2 Drs.Sci, or 1 Dr.Sci and 1 Cand. Sci., who will study 
the manuscript in details and prepare reports. For defense of Dr. Sci. 
thesis, the Special Council selects three reviewers (3 Dr.Sci.) with 
a backgroud in a field relevant to the defended thesis. In both cases 
one astronomical institution in the relevant scientific field is appointed 
to be an independent ``Leading Referee Institution''. The signed reviews 
must be approved by the Director of this institution.

In the third step, the Supreme Attestation Commission of Ukraine 
approves/does not approve this thesis as well as awards/does not 
award the Cand.Sci. and Dr.Sci. degree. The Supreme Attestation 
Commission (SAC) of Ukraine (www.vak.org.ua) is a government 
institution, which develops the general rules for processing thesis 
manuscripts, thesis defense and approval as well as awards 
the scientific degrees. This institution also approves the 
membership of the Special Councils on Thesis Defense. The Special 
Expert Councils under SAC were established to consider possible 
conflict situations. Members of these special councils must have 
a Dr. Sci. degree in a specialty relevant to that of the defended 
thesis. Membership in these councils is renewed once every three years. 

In the current system a positive defense at the Special Council 
meeting is possible even when a negative report is prepared by the 
reviewers. The Special Council may also decide that the Cand. Sci. 
thesis satisfies requirements for the Dr. Sci. thesis and can nominate 
the candidate for the higher Dr. Sci. Degree. In case of a conflict 
situation the SAC Expert Council has the right to reverse the decision 
of the Special Council on Thesis Defense 
following established procedures. It also is possibile to 
prepare and to defend a thesis in an interdisciplinary field, e.g., 
astronomical instrumentation, which is more related to the SAC 
requirements for theses in Technical Sciences, history of astronomy 
or methodology of astronomy education.  

At present, the Special Council on Thesis Defense which operates at 
the Main Astronomical Observatory of the NASU in Kyiv, nominates astronomers 
for Dr. Sci. and Cand. Sci. in Phys.\& Math. It covers all the SAC adopted 
specialties for astronomers: ``Heliophysics and Physics of Solar System'', 
``Astrophysics and Radio Astronomy'', ``Astrometry and Celestial Mechanics'', 
and ``Methods of  Remote Sensing of the Earth''. The Special Council 
operated in I.I. Mechnikov National University of Odesa nominates for 
Cand. Sci. degree in the fields of: ``Theoretical physics'', ``Astrophysics 
and Radio Astromomy''. The Special Council of the V.N. Karazin 
National University of Kharkiv nominates for Dr. Sci. and Cand. Sci. 
degrees in ``Astrophysics and Radio Astromomy''.

Since 1992, about 130 astronomers were awarded higher scientific 
degrees (20\% of them Dr. Sci. Degree). Our present-day problem is a 
brain drain of young scientists: currently 50\% of those who obtained 
a Cand. Sci. degree work outside of Ukraine. 

\subsubsection{Scientific Careers.} 
The nomenclature of scientific positions 
is the following: \\ 
{\em Junior Staff Scientist} -- holders of a Master's degree 
 who are starting the scientific activity. \\
{\em Research Staff Scientist}-- scientists who defended 
Cand. Sci. thesis. \\ 
{\em Senior Staff Scientist} -- those who worked successfully 
for at least five years as Research Staff  Scientist. \\ 
{\em Leading Staff Scientist} -- scientists who defended Dr. 
Sci. Thesis. \\
{\em Principal Staff Scientist} -- senior scientist who 
worked successfully at the position of Leading Staff Scientist. \\

The nomenclature of faculty positions in higher educational 
institutions is the same as in many countries: Lecturer, Senior Lecturer, 
Assistant-Professor, Professor.

Unfortunately, for the time being, the Soviet type system of long-term 
fixed staff positions has been preserved without substantial changes. 
There is little competition for permanent positions. In fact, any young 
scientist can get a position at an astronomical institution practically 
for the rest of his or her life. Post doctoral positions are not part of 
the current structure of scientific positions.

The following academic titles/status 
are used for scientists: \\ 
{\em Senior Researcher} -- those who defended Cand. Sci. thesis and worked 
successfully at least two years at the position of Senior Staff Scientist \\ 
{\em Assistant-Professor} -- those who defended Cand. Sci. thesis and worked 
at least three years at the position of Assistant-Professor \\ 
{\em Professor} -- Assistant-Professors who defended the Dr. 
Sci. Thesis, and Senior Researchers who defended the Dr. Sci. thesis 
and were supervisors of at least five Candidates of Science \\
{\em Corresponding Member of the National Academy of Sciences of Ukraine} 
-- senior scientists \\
{\em Member of the National Academy of Sciences of Ukraine} 
-- advanced senior scientists.

\subsection{Government and Non-government Astronomical Institutions}

\subsubsection{Organization of Scientific Activity in Astronomical Institutions} 
All questions and decisions concerning scientific activities 
(research programs, scientific careers of employees, nominations 
for academic status, final reports for various projects, scientific 
theses research, structure and staff of departments and laboratories 
etc.) must be considered and approved by the Scientific Council of 
an institution. The Council is formed by the Director of the institution 
and consists of the senior, leading and principle staff scientists. 
Membership in this council must be approved by a majority vote of 
scientists of the institution. Each member of the Council has one vote. 
A majority vote is sufficient to approve most decisions, however, 
nominations for staff positions and scientific titles still require a 
2/3 majority of total members. 

Research programs can be divided into two types, depending on the 
sources of funding: government budget programs and off-budget programs 
(see sections  3.6 and 4 for details). The Director, heads of departments, 
and program managers form teams of scientists and engineers to 
conduct the research programs. Structure of these teams may not follow a 
formal distribution of staff by departments or laboratories. Still, the 
project manager, pricipal investigator (PI) and the main department 
responsible for the project are always known. As a rule government funded 
programs are planned for 3 - 5 years. After completing the project the team 
members write a final report, which must be refereed by another institution. 
The final report must then be presented and approved at a Scientific Council 
meeting. All decisions of the institution's Scientific Council must be 
approved by the Scientific Council or Scientific Bureau of a higher 
governing body.

The importance of the role of Director, Scientific Council, 
heads of departments and project managers for scientific activity of the 
institution depends on many subjective and objective factors and may vary 
from institution to institution.

Unfortunately, under current research bugetary conditions the principle 
``what  is possible'' often prevails over ``what is more interesting''.
 The heads of departments and research project managers are only the 
 formal owners of the project's money. Due to practical reasons, 
 to minimize possible problems and loses, usually one person (as a rule, 
 the Director of the institution) is responsible for resolving all financial 
 issues. 

\subsubsection{Governmental Astronomical Institutions} 
In general, 12 governmental 
institutions are engaged in astronomical scientific activity. They 
are governed by two different higher government level organizations: 
the National Academy of Sciences of Ukraine (NASU) and the Ministry of 
Education and Science of Ukraine (MESU). 

As mentioned in Section 2, the largest astronomical institutions 
are established within the structure of the NASU (www.nas.gov.ua). 
NASU is a self-governing organization whose activity is aimed at S\&T, 
social-economical, and cultural development of Ukraine. 
The Government of Ukraine (www.kmu.gov.ua) forms the NASU budget. 
In its own turn, the NASU Presidium, through relevant scientific NASU 
Divisions, forms a budget of all subordinated institutions. The role 
of the NASU in the national basic and applied S\&T policy is the same 
as the Max Plank Geselschaft in Germany or the Centre National de 
Recherche Scientifique in France.

The following four astronomical institutions are subordinated to the NASU: \\
$\bullet$ Main Astronomical Observatory (www.mao.kiev.ua ) in Kyiv \\
$\bullet$ Institute of Radio Astronomy (www.ira.kharkov.ua ) in Kharkiv \\
$\bullet$ Poltava Gravimetric Observatory (pgo@poltava.ukrtel.net ) of the Institute of Geophysics \\
$\bullet$ International Center for Astronomical \& Medical-Ecological  Research (ICAMER), in Terskol, North Caucasus, Russia 
(www.mao.kiev.ua/icamer or (www.allthesky.com/observatories/terskol.html). 

The first two institutions are related to the NASU Division for 
Physics and Astronomy, the third one to the NASU Division for Earth 
Sciences. The bureaus of these divisions approve decisions of 
Scientific Councils of the mentioned astronomical institutions. 
The ICAMER is an interdivisional institution and its activity is 
regulated in the framework of a special memorandum signed by the 
NASU and the Russian Academy of Sciences and by the Governments of 
Ukraine and Russia.  

The Ministry of Education and Sciences of Ukraine (www.mon.gov.ua) 
governs seven institutions engaged in astronomical research. 
Two of them are directly subordinate to the MESU: \\ 
$\bullet$	Scientific Research Institute, Crimean Astrophysical 
Observatory,\\  
(www.crao.crimea.ua ) in Naukove, Crimea \\
$\bullet$	Research Institute ``Mykolaiv Astronomical Obsevatory'' \\
 (www.mao.nikolaev.ua)  in Mykolaiv. 

Three astronomical observatories function 
as Research Institutes at universities: \\
$\bullet$ Scientific Research Institute ``Astronomical Observatory'' of the 
I.I.~Mechnikov 
National University of Odesa (astro@paco.odessa.ua) \\
$\bullet$ Institute of Astronomy of the Karazin National University of 
Kharkiv (www-astron.univer.kharkov.ua) \\
$\bullet$ Astronomical Observatory of the Ivan Franko National 
University of L'viv (www.astro.franko.lviv.ua ). 

Two small astronomical institutions function as departments of universities: \\ 
$\bullet$ Space Research Laboratory, National University of Uzhgorod \\  
(www.univ.uzhgorod.ua/nauka/zsurnal/Lab-space.html), \\
$\bullet$ N.I.Kalinenkov Astronomical Observatory, Pedagogical State 
University of Mykolaiv (office@mdpu.edu.ua). 

These universities allocate the budgets of the observatories as 
well as maintain Departments of Astronomy for training students. 
The Scientific Councils of universities approve all decisions of 
the Scientific Councils of the observatories.

The Shevchenko National University of Kyiv is a self-governing 
institution and its budget is determined by the central government. 
An Astronomical Observatory functions there as a Research Laboratory 
of the Department of Astronomy and Space Physics of the Physics Faculty \\
(www.observ.univ.kiev.ua ).

Information on staff membership and research fields of larger 
astronomical institutions is presented in Table 1 (Vavilova \& Yatskiv 2003). 


\begin{table}
\begin{center}
\caption{Staff membership and research fields of larger astronomical 
institutions in Ukraine}
\begin{tabular}{lll}
\hline

Institution & Total staff/    & Recearch \\
            &  Scientific staff/ &      fields \\
            & Cand and Dr. Sci. &             \\

\hline

Main Astronomical   & 213/90/69 &  Extragalactic Astronomy \\
Observatory          &            &  Physics of stars \& brown dwarfs \\
of the NAS of Ukraine        && Positional Astronomy \\
        & &                                Solar System Bodies \\
       & &                                Solar Physics \\
        & &                                Space Geo-dynamics \\
        & &                                Space Plasma Physics \\

Scientific Research Institute           & 358/92/58 & Extragalactic Astronomy\\
``Crimean Astrophysical  &        & Ground-based and \\
 Observatory''                    &        &  Space-born Instrumentation\\
of the MES of Ukraine                                                &&  Radio Astronomy (mm, cm) \\
                                        &      &  High-energy astrophysics\\
                                        &        & Physics of stars \\
                                        &       &  Solar System Small Bodies\\
                                        &        &  Solar Physics, Solar Activity \\

Institute of Radio Astronomy & 306/102/88 & Radio astronomy\\
of the NAS of Ukraine                  && (decameter and mm) \\

Astronomical Observatory of       & 64/35/26 &        Astrometry \\
the Shevchenko National         &&                     General Relativity \\
University of Kyiv              &&              Extragalactic Astronomy \\
                                &    & Solar Physics, Solar Activity \\
     &&     Solar System Small Bodies \\

Astronomical Observatory & 28/16/12   & Extragalactic Astronomy \\
of the Ivan Franko National     && Cosmology \\
University of L'viv      && Satellite Geodesy \\
                         && Solar Physics, Solar Activity \\

Scientific Research Institute         & 75/65/26  & Physics of the Solar System \\
`` Astronomical Observatory''       &&                  Small Bodies \\
of the I.Mechnikov         &  & Variable stars \\
National University of Odesa  & &Physics of stars \\

Institute of Astronomy of            &  83/43/20 &  Ground-based Instrumentation \\
V. Karazin National          &                   & Physics of stars \\
University of Kharkiv                          &                   & Solar activity \\
                                     &                   & Solar System Small Bodies \\

Mykolaiv Astronomical   & 75/19/10     & Ground-based Instrumentation \\
Observatory             &     & Positional astronomy \\
of the MES of Ukraine   && \\

\hline
\hline
\end{tabular}
\end{center}
\end{table}

Taking into account quantitative factors, {\em i.e.~} the number of scientists 
engaged in astronomical research per population, the number of 
astronomical institutions as well as astronomical 
infrastructure (see Section 3.4), we could consider Ukraine a 
large astronomical 
country in Europe. In total, more than twenty observatories 
and departments at various scientific institutions and universities are 
engaged in astronomical research. As to the qualitative factors, i.e. 
number of publications in world recognized journals, citation index etc., 
the situation is not so clear. As can be seen in Table 1, research at 
Ukrainian observatories covers a wide range of disciplines. In the case of 
observational programs, access to modern astronomical facilities is rather 
limited. As a result in many cases theoretical interpretations of 
observations conducted at other observatories still prevails. On the other 
hand, Ukraine has developed its own astronomical infrastructure (see 
section 3.4). However these astronomical facilities need to be upgraded 
rapidly to provide any competition for leading observatories of the world.

\subsubsection{Non-governmental Astronomical Institutions} Since 1991 the 
Ukrainian 
Astronomical Association, UAA, \\ 
(www.observ.univ.kiev.ua/uaa) has coordinated 
astronomical activity in Ukraine. The UAA consists of 16 Institutional Members 
and dozens of Individual Members, with a total membership of about 1500 
astronomers. From 1992 the UAA serves as the National Committee of the 
International Astronomical Union, IAU, and as an Affiliated Society of the 
European Astronomical Society, EAS. One hundred and sixty two Ukrainian 
astronomers are IAU members and 96 are EAS members.

There are a few other non-governmental institutions related to astronomy. 
Two of them, Odesa Astronomical Society and the Ukrainian Society of 
Gravitation, Relativistic Astrophysics and Cosmology, are UAA Associate 
Members. Two years ago the Ukrainian Society of Amateurs of Astronomy was 
founded under the patronage of professional astronomers.

\begin{figure}
\psfig {file=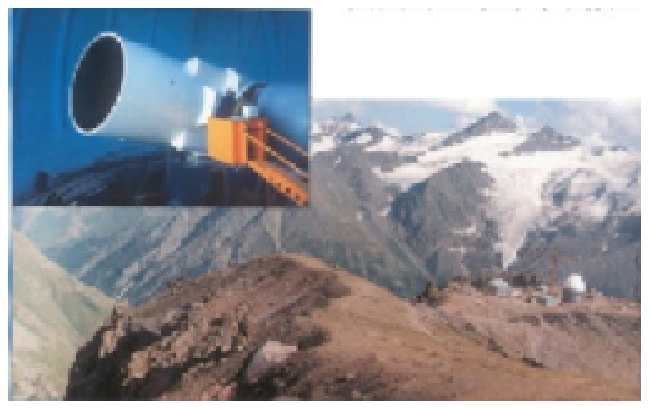,height=4cm} 
\psfig {file=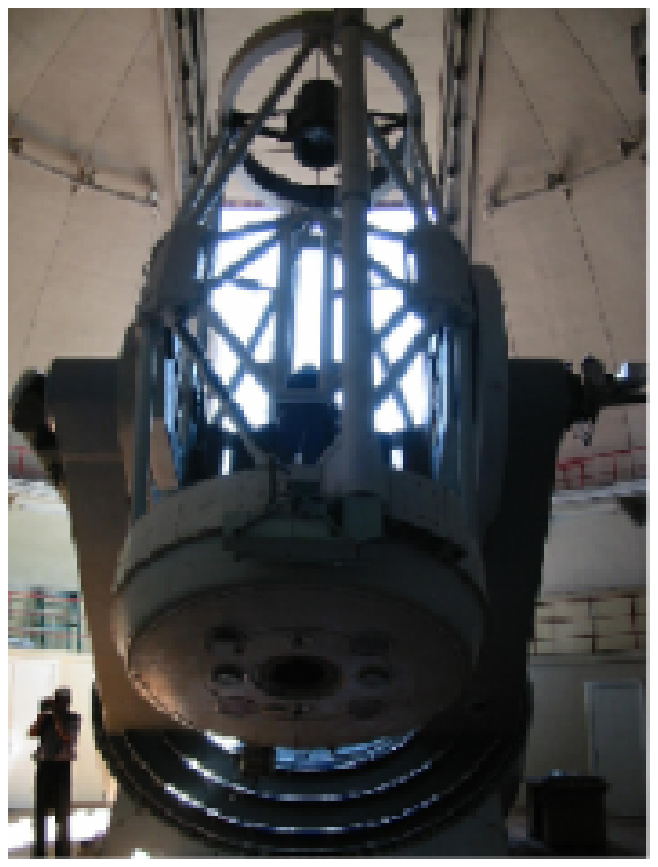,height=4cm}
\psfig {file=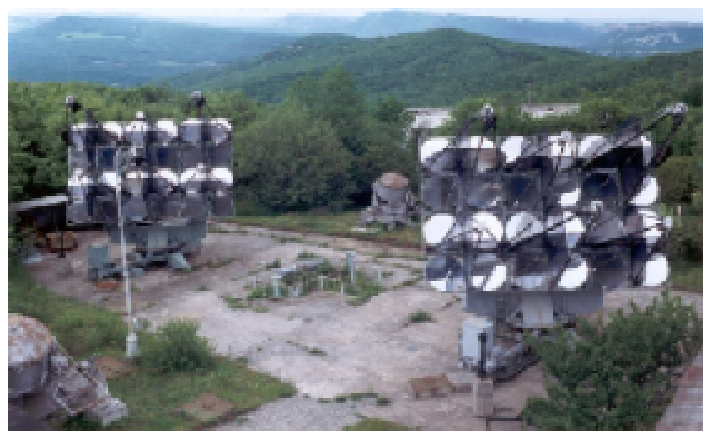,height=4cm}
\psfig {file=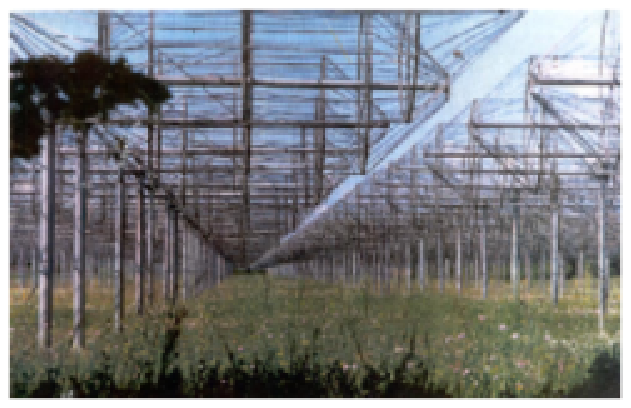,height=42mm}
\caption[]{\label{_NTIO_} From top: the 2 m telescope of ICAMER (Peak Terskol, 
North Caucasus)and Acad. Shajn 2.6 m telescope of Crimean Astrophysical 
Observatory (Naukove, Crimea). Gamma Telescope  GT-48 of 
Crimean Astrophysical Observatory and Decameter Radio Telescope UTR-2 
(Kharkiv), see more on www.mao.kiev.ua/staff/yp/OSA7/a.htm.
Copyright 2003 Ukrainian Astronomical Association (1,3,4) and 
A.V.Terebizh(2).}
\end{figure}

\subsection{Infrastructure: Large, Moderate, and Small Telescopes and 
Telescope Networks.}

Ukrainian astronomical institutions possess a wide range of telescopes. 
Many of them were constructed 30 and more years ago, but are still in use. 
The main problem is to upgrade these telescopes and equip them with  
modern detectors and other devices for making observations and obtaining 
results of sufficient quality.

\subsubsection{Largest Telescopes and Networks, (see also Fig. 2)}

\begin{itemize}
\item	The UTR-2, Ukrainian T-shape Radio telescope,  
	(www.ira.kharkov.ua/ utr2), is the largest array in the world 
	operated at the decameter wavelengths, extrememly low 
	frequencies <25 MHz. This telescope belongs to the Institute of 
	Radio Astronomy, IRA, of the NASU. It is located near Grakovo 
	village, about 80 km from Kharkiv (northeastern Ukraine). 
	The effective area of the UTR-2 (152 000 sq. m) is more than the 
	effective area of all existing radio astronomical telescopes 
	put together. The resolution is about of 40'x40' at the middle 
	frequency, 16.7 MHz. 

\item	A decametric Very Long Baseline Interferometry, VLBI, network 
	URAN was built with the UTR-2 as the basis. Besides UTR-2 it 
	consists of four additional radio telescopes with sizes 5 to 10 
	times less than that of UTR-2: URAN-1 near Kharkiv, 
	URAN-2 near Poltava, URAN-3 near L'viv, and URAN-4 near Odesa. 
	They are electrically phased steering arrays operating from 
	10 to 30 MHz. Baselines from 40 km to 900 km provide an angular 
	resolution from several minutes to one second of arc. The angular 
	resolution of 1 arcsec corresponds to the fundamental limit 
	imposed by scattering at these frequencies in the interstellar 
	medium. These network telescopes also belong to the Institute of 
	Radio Astronomy.

\item	The 70-meter dish radio telescope RT-70 is a highly efficient fully 
	tracking instrument located near Evpatoria in Crimea. Its effective 
	area is of 2500 sq. m  and its beam width is 2.5 arcmin at the 5-cm 
	radio wave length. There are only 10 antennas of such size in the 
	world. This telescope is being up-graded to provide astronomical 
	research at wavelengths 92, 18.6, and 1.35 cm for future work in 
	the European VLBI network (EVN). This antenna belongs to the 
	National Space Agency of Ukraine, NSAU.

\item	The RT-22 is a precise radio telescope operating at mm and cm 
	radio wavelengths located in Simeiz, Crimea. It has a Cassegrain and 
	prime focus feed system on an azimuth-elevation mount and its 
	characteristics are: diameter 22 m, surface tolerance 
	(root mean square) 0.25 mm, wavelength limit 2 mm, and focal 
	length 9.525 m. RT-22 is included in VLBI astrophysiscal and 
	geodetic projects with the European and USA networks. 
	This instrument belongs to the SRI ``Crimean Astrophysical 
	Observatory'' of the MESU.   

\item	The Acadimician Shajn  2.6-meter reflector is the largest optical 
	telescope in Ukraine. The  telescope was built in 1961. Its 
	equatorial mount supports a 2.6-m parabolic primary with several 
	optical systems: primary (F/4 and with a focal reducer F/2.6), 
	Cassegrain (f/16), Nasmith (f/16), and two f/40 coude foci, direct 
	and bent.

\item	2-meter ZEISS telescope at the Peak Terskol, North Caucasus, 
	Russia (www.allthesky.com/observatories/terskol.html). It belongs 
	to the ICAMER.
\end{itemize}

\subsubsection{Moderate-size Telescopes}.

\begin{itemize}

\item	AZT-11 (Crimean Astrophysical Observatory) is 1.25-m Ritchey-Cretien 
	reflector, built in 1981. Focus length is 16 m, available foci are  
	Main Cassegrain and Auxiliary Cassegrain. An offset photoelectric 
	auto guider is provided for the Main focus. A TV guider with a 30-cm 
	refractor and a 40 arcmin field-of-view also is available. Objects 
	brighter then 15 mag can be resolved. A computer based control system 
	provides automated pointing with 15 arcsec precision and other 
	services, i.e. fine tracking of fast moving objects 
	(comets, asteroids), access to object catalogues, and dome-telescope 
	synchronization.

\item	The Tower Solar Telescope TST-1 (Crimean Astrophysical Observatory). 
	A 120 cm coelostat and a 90 cm spherical primary mirror feed the 
	telescope to provide f/56 or f/78 Cassegrain foci equipped with 
	spectrographs.

\item	ACU-5 (Main Astronomical Observatory of the NASU, \\
	www.mao.kiev.ua/sol/sol\_w1.html) consists of a 440-mm coelostat 
	and an additional mirror, 440/17500-mm main mirror and 200-mm 
	Cassegrain mirror system with a 60-m equivalent focal length. 
	The spectrograph camera and collimator mirrors are made out of 
	one single block of glass of 500/7000-mm, the grating has a ruled 
	area of 140x150 mm with 600 lines per mm. 

\item	Solar ACU-26 telescope (MAO NASU, www.mao.kiev.ua/sol/sol\_w2. \\
        html) 
	was constructed at the Peak Terskol in 1989. The diameter of the main 
	spherical mirror is 650 mm with a focal length of 17750 mm. 
	The telescope is equipped with a 5-camera spectrograph permitting 
	simultaneous observations in five spectral regions. The diameter of 
	the collimator and cameras is 300 mm, the focal length 8000 mm. 
	The 250 mm x 200 mm grating, 600 lines/mm, permits dispersion 
	in fourth order of 21.9 mm/nm at 395.0 nm and 33.0 mm/nm at 650.0 nm.

\item	SLR (MAO, NASU, Crimean Laser Observatory in Katsiveli, Crimea) 
	is a 100-cm telescope with Ritchey-Cretien and Coude systems on an 
	English mounting. The equivalent focal length of Ritchey-Cretien 
	system is 13.3 m and of the Coude system is 36.5 m. A CCD-camera 
	with 256 x 256 pixels allows positional and photometric observations. 
	A satellite ranging laser is mounted at the Coude focus. The Crimean 
	satellite laser ranging, SLR, Station N1873 is a member of a SLR world 
	network and participates in the majority of international programs 
	observing satellites. 

\item	The gamma ray telescope GT-48 (Crimean Astrophysical Observatory) 
	is designed for searching and investigating sources of very high 
	energy (VHE) gamma radiation (~1012 eV) by measuring Cherenkov 
	flashes in the Earth atmosphere on moonless nights. The installation 
	GT--48 consist of two independent alt-azimuth arrays 20 m apart. 
	Each array consists of six 1.2-m telescopes with a common focus. 
	Three of them are designed for detection of short ultraviolet 
	Cherenkov radiation initiated by cosmic radiation, gamma-rays 
	as well as charged particles, and have solar blind photomultipliers 
	in their focal planes. The other 3 telescopes image the flashes with 
	37 photomultipliers (imaging camera).

\item	The 1.24 m Ritchey-Cretien reflector (Simeiz, Crimean Astrophysical 
	Observatory). Diameter of the second mirror is 0.35 m and the focal 
	length is 14.5 m on an English mounting (EM-2). A synchronous driving 
	gear is used with a quartz stabilizer for guiding.  
\end{itemize}

\subsubsection{Small Telescopes.} 

Small-size telescopes are listed in Table 2.

\begin{table}
\begin{center}
\caption{Small-size telescopes of Ukraine }
\begin{tabular}{ll}
\hline

Main Astronomical Observatory,      & AZT-2 (80cm) \\
NAS of Ukraine                      & GPS station (1 m) \\
                                    & Axial Meridian Circle \\
                                    & Twin astrograph (0.7 m) \\
				    & 2 GPS stations (1 m) \\
                                    &                         \\
Crimean Astrophysical Observatory,  & AZT(0.5 m) \\
MES of Ukraine                      & MTM-500 (0.5 m) \\
                                                        & AZT-8 (0.7 m) \\
                                    &                                  \\
 Astronomical Observatory, &     AZT-8 (0.7 m), AZT-14 (0.5 m) \\
 Shevchenko National       &   Twin astrograph (0.4 m)                  \\
 University of Kyiv        &   Horizontal Solar Telescope (0.8 m)       \\
                           &                                            \\
Astronomical Observatory,            & 1 m and 0.6 m telescopes \\
I.I.Mechnikov National               & two 0.8 m telescopes\\
University of Odesa                  & two 0.5 m telescopes\\
                                     &                      \\
Institute of Astronomy,              & AZT-8 (0.7 m) \\
V.N.Karazin National                 &             \\
University of Kharkiv                &             \\
                                     &                \\
Mykolaiv Astronomical Observatory,   & Axial Meridian Circle  \\
MES of Ukraine                       & Multi-Channel Telescope \\
                                     & GPS station (1 m) \\
                                     &                      \\
Astronomical Observatory,                 &     AZT-2 (0.8 m) \\
I.Franko National                    & \\
University of L'viv               &  \\
                                  &   \\
Space Research Laboratory         &     SPL-telescope (1 m) \\
Uzhgorod National University      &                        \\
                                  &                         \\
Narodna Observatory               &     60-cm ZEISS telescope \\
Andryushivka, Zhytomir region     &                           \\
				  
\hline
\hline
\end{tabular}
\end{center}
\end{table}

\subsection{Astronomical Publications}

Ukrainian astronomers prefer to publish their work in international 
journals such as ``Astrophysical Journal'', ``Astronomy and Astrophysics'', 
``Astronomical Journal'', 
``Solar Physics'', ``Icarus'', etc. Indeed, such publications provide 
the opportunity to publicize their work to a world-wide 
astronomical community. A few refereed journals, which publish original 
papers on astronomical research are published in Ukraine: 

\begin{itemize}

\item ``Kinematics and Physics of Celestial Bodies'' (in Russian and Ukrainian, 
since 1985, bimonthly) covers various fields of modern astronomy. The 
English translated version is available from Allerton Press, New York.
``Space Science and Technology'' (in Russian and Ukrainian, since 1995, 
issued quarterly) contains papers on space astronomy and physics.

The journals ``Kinematics and Physics of Celestial Bodies'' and 
``Space Science and Technolgy''  
 are published by the publishing department of the Main Astronomical 
 Observatory of the NASU \\ (www.mao.kiev.ua/eng/papers\_e.html).

\item ``Radio Physics and Radio Astronomy'' (in Russian and Ukrainian, 
since 1995) is a quarterly journal published by the Institute of 
Radio Astronomy of the NASU. It covers various problems on formation, 
propogation and registration of radio waves in different media.

\item ``Bulletin of the Crimean Astrophysical Observatory'' has been published 
by this observatory since 1947. Ninety-nine volumes have already been 
published. Starting with volume 57 published in 1977, the ``Bulletin of the 
Crimean Astrophysical Observatory'' is translated into English and 
distributed by the Allerton Press, New York.

\item ``Odessa Observatory Publications'' (in English, 2 issues per year) 
contains publications of original astronomical research and papers 
presented at international conferences.

\item  The annual ``Bulletin of the Shevchenko National University of Kyiv. 
Astronomy'' contains results of research conducted by astronomers 
from the University observatory.

\item The ``Information Bulletin of the Ukrainian Astronomical Association'' 
is published twice per year and contains information about current UAA 
activity as well as proceedings of UAA meetings.

\end{itemize}

Besides these scientific journals, the Main Astronomical Observatory has 
been publishing ``The Astronomical Calendar'' since 1996  (``The Short 
Astronomical Calendar'' in 1948-1995), devoted to disseminating astronomical 
knowledge for amateurs and students. Another astronomical calender, 
``The Odessa Astronomical Calendar'' is published by the Odesa Astronomical 
Society.

Journals such as ``Our Sky'' (since 1998, Kyiv Republican Planetarium), 
``Universe: Space and Time'' (since 2003, private issue), ``Pulsar'' 
(since 1998, Association of Trade Union Organizations of Students of Kyiv), 
as well as the Ukrainian version of the translated ``Scientific American'' 
containing many papers for nonspecialists also circulate in the popular 
scientific media.

\section{Current Investments in Astronomical Research}

The budgetary investments in astronomy are made by different 
governmental agencies. The Government of Ukraine provides through 
the NASU and MESU (see section 3.3) the main funding of astronomical 
institutions. It covers salaries of staff scientists, engineers, and other 
personnel as well as overhead for maintenance work on government buildings 
in institutions. To a lesser degree it covers expenditures for equipment, 
travel, and supplies.

About 30 \% funding is obtained from orders from the National Space Agency 
of Ukraine (www.nkau.gov.ua/nsau) and from other government institutions, 
for example, the State Fund for Basic Research (www.dffd.gov.ua). Another 
10 \% of total funding is from individual and collaborative international 
grants under research programs of NATO, INTAS, STCU, CRDF, UNESCO, USAID. 
Besides additional salaries for employees of such collaborative projects 
these grants provide funding for equipment, travel, and system network 
development. 

Total expenditures of the largest observatories are at least 2 000 000 USD, 
smaller observatories spent about of 350 000 USD per year.

\section{International Cooperation}

There are at least two forms of collaboration with  
foreign colleagues:

Participation in international projects in the form of 
individual grants or projects, as collaborators. 
As a rule, coordinators or prinicipal investigators are 
foreign scientists. Ukrainian participants obtain some 
financial support for short time periods (up to three months) 
and travel abroad. 

Collaborative projects are conducted within the framework of 
international research programs. 
A few of current projects are listed here: \\
-- Civilian Research and Development Foundation, CRDF, 
project: ``Spectroscopic and Photometric Monitoring of Selected 
Active Galactic Nuclei Objects with Extreme Properties'' (Peterson, 
B. M., USA and Pronik, V. I., CrAO, Ukraine). \\
-- International Association for the Promotion of Co-operation with 
Scientists from the New Independent States of the Former Soviet Union,
 INTAS, project: ``The Investigation of Young Stars with 
Protoplanetary Disks'' (Grinin V. P., CrAO). \\
-- CRDF project: ``High-energy gamma-quanta investigation with ground-based 
Cherenkov telescopes''  (Fomin, V.P., CrAO). \\
-- INTAS project: VLBI - astrophysical and geodetic projects 
with European and USA Networks (Volvatch, A.E., CrAO). \\
-- Science and Technology Center of Ukraine, STCU, project: ``Fundamental 
Physics and Astrophysics on-board the International Space Station: 
Theoretical Basis for General Relativity Tests and Astronomical 
Support of the Asteroid-Hazard Observational 
Program'' (Vavilova, I.B., UAA).

Some scientists obtain international grants for long term work 
at foreign observatories or astronomical institutions. Due to 
internal rules and to protect staff positions, the foreign 
tenure cannot be longer than 5 months. 

Several groups participate in joint projects with their own 
telescopes, hardware, know-how, finances and staff. This form 
of international participation is the most challenging.

In recent years significant contributions to world astronomy were made 
by: Y. Izotov, N. Guseva, L. Pilyugin, and V. Karachentseva 
(extragalactic astronomy); Y. Shkuratov, D. Lupishko, and V. Rozenbush 
(physics of planets and solar system small bodies); L. Shulman and K. 
Churyumov (cometary physics); Y. Yatskiv (nutation and reference frames, 
member of the team awarded the Descartes Prize of the European Union 
in 2004); R. Kostyk, N. Shchukina, N. Stepanyan, and V. Kotov 
(solar physics and solar-terrestrial relationship); V. Pronik and I. 
Pronik (AGNs); R. Hershberg, L. Lyubimkov, V. Andrievsky, and I. 
Andronov (physics of variable stars); A. Konovalenko (decameter radio 
astronomy); N. Steshenko (astronomical instrumentation). 
Most of their prime results were obtained due to the tight international 
cooperation.     

\section{Scientific Investment and Priorities} 

\subsection{Scientific Priorities}  Though it is a very difficult task to 
compose a realistic research planning document taking into account the 
economic situation in Ukraine, the scientific and investment priorities for 
Ukrainian astronomy were specified after discussions at the UAA Meeting in 
2003: \\
$\bullet$ Formation and evolution of galaxies \\
$\bullet$ 	Global characteristics of the Sun and Sun-like-stars \\
$\bullet$ 	Ground-based support of space missions \\
$\bullet$ 	Observational and theoretical cosmology  \\
$\bullet$ 	Physics and kinematics of solar system bodies \\
$\bullet$ 	Solar-planetary interactions 

Understanding astronomical phenomena requires high-quality data in 
all frequency ranges. The activity of Ukrainian astronomical institutions 
is concentrated on ground-based observations in optical, centimeter and 
decameter wavelengths of the electromagnetic spectrum. Space mission 
data over a wide range of wavelengths are also used. 

It is useful to distinguish short-term and 
medium-term investment priorities of Ukrainian astronomy.

\subsubsection{Short-term Investment Priorities} Astronomical facilities, 
which are part of international networks or are involved in 
conducting international programs must be upgraded, in particular: \\
-- The RT-22 radio telescope of the Crimean Astrophysical Observatory 
must be equipped with the new MARK4 or MARK5 recording systems and a 
new hydrogen frequency standard. \\
-- R\&D for up-grading the UTR-2 decameter telescope of the Institute 
of Radio Astronomy must be finished and a new type of decameter antenna 
has to be tested. According to the decision of the NASU this work has 
to be done in 3 years. In the future Ukraine would like to be involved 
in the Square Kilometre Array, SKA, project. \\
-- The optical 2.6-m telescope of the Crimean Astrophysical Observatory 
must be equipped with the same Echelle-grating spectrometer as the 
2-m telescope at the high altitude observatory at Terskol Peak and 
with a multicolor photometer-polarimeter. \\
-- The optical telescopes, which are used for observations 
of solar system small bodies, stars, and near Earth objects, NEOs, 
have to be equipped with new CCD-cameras and computer controlled systems.

Investments will be also allocated for establishing ground-based 
networks of small-size and medium-size optical telescopes, which 
are or will be involved in international monitoring projects for 
studying gravitational microlensing; ultra-rapid variability of stellar 
brightness and polarization; multi-wave monitoring of red dwarf flare 
stars and cataclysmic variable stars; magnetic activity of the Sun and 
solar-like stars; and pulsating stars as single objects vs components 
of multiple systems.

\subsubsection{Medium-term investment priorities} 
As to the medium-term priorities, 
the attention will be paid to participation of Ukrainian astronomical 
institutions in space missions, e.g., Spectrum-Radio Astron (2006), 
Spectrum-UV (WSO) (2008); preparation of the Solar-Oriented Telescope 
(SOT) in the framework of the International Living with a Star Program; 
and a prospective Lunar space mission. Some projects are envisaged for 
the International Heliophysical Year.

Special attention will be paid to the development of the 
Ukrainian Virtual Observatory as a part of the International 
Virtual Observatory Alliance. 
For example, the work concerning estabvishment of the Virtual X-ray 
and Gamma Observatory (VIRGO) as Ukrainian part of the INTEGRAL project 
has been starting in 2005.

At the moment new large ground-based 
astronomical facilities are not foreseen in Ukraine. 

\section{Conclusion}

Despite all problems the future of Ukrainian astronomy can be 
considered promising. Traditionally, Ukraine is a country with a very 
high level 
of education and culture, and basic research remains an important part 
of science in Ukraine. Existing observational facilities provide unique 
opportunities for the study of astronomical objects in a wide range of 
spectral regions and new generations of astronomers can use these national 
facilities. Ukraine is one of a few counties developing its own space 
technologies, launch vehicles and programs. Most of these space projects 
are conducted with wider international cooperation. 

Globalization of science provides new opportunities for collaboration 
with astronomers of other countries. The main goal of Ukrainian astronomy 
is active participation in the development and use of large infrastructures 
of the international astronomical community. However, this can be achieved 
only if  adequate support can be obtained from the government. 

Astronomy as well as any field of scientific research is 
the innovative force of economic development. Astronomers are 
a unique tightly knit community of people dedicated to science, 
trying to understand the nature of the universe. Ukrainian astronomers 
are in the mainstream of world science and they have to maintain their 
place within it.

\section{Acknowledgements}

We thank our colleagues Shakovskaya, N. (Crimea), Stodilka, M. (Lviv), 
Andronov, I. (Odesa), Tsymbal, V. (Simpheropol), for their help and 
information. We thank Yatskiv, Y. (Kyiv) for helpful remarks and 
enlightening discussions. YP's studies were partially supported by 
PPARC, Royal Society, Leverhulme Trust. 
I.V. thanks the Oxford University for hospitality during the completion 
of the final version of this paper.

\end{document}